\documentclass{pasj01}

\draft

\begin{document}
\SetRunningHead{S. Takeda et al.}{Running Head}

\title{Suzaku observations of the hard X-ray spectrum of Vela Jr.\ (SNR RX J0852.0$-$4622)}


 \author{%
   Sawako \textsc{Takeda},\altaffilmark{1}
   Aya \textsc{Bamba},\altaffilmark{2}
   Yukikatsu \textsc{Terada},\altaffilmark{1}
   Makoto S. \textsc{Tashiro},\altaffilmark{1}
   Satoru \textsc{Katsuda},\altaffilmark{3}
   Ryo \textsc{Yamazaki},\altaffilmark{2}
   Yutaka \textsc{Ohira},\altaffilmark{2}
   and
   Wataru \textsc{Iwakiri}\altaffilmark{4}}
 \altaffiltext{1}{Graduate School of Science and Engineering, Saitama University, Shimo-Okubo 255, Sakura, Saitama 338-8570}
 \email{takeda@heal.phy.saitama-u.ac.jp}
 \altaffiltext{2}{Graduate School of Science and Engineering, Aoyama-Gakuin University, 5-10-1 Fuchinobe, Chuo, Sagamihara 252-5258}
 \altaffiltext{3}{Department of High Energy Astrophysics, Institute of Space and Astronomical Science (ISAS),
 Japan Aerospace Exploration Agency (JAXA), 3-1-1 Yoshinodai, Sagamihara 229-8510}
 \altaffiltext{4}{RIKEN (The Institute of Physical and Chemical Research) Nishina Center, 2-1 Hirosawa, Wako 351-0198}


\maketitle

\begin{abstract}
We report the results of Suzaku observations  of the young supernova remnant, Vela Jr.\ (RX J0852.0$-$4622), 
which is known to emit synchrotron X-rays, as well as TeV gamma-rays.
Utilizing 39 Suzaku mapping observation data from Vela Jr., 
a significant hard X-ray emission  is detected with 
the hard X-ray detector (HXD) from the north-west TeV-emitting region.
The  X-ray spectrum  
is well reproduced by a single power-law model with the photon index of 3.15$^{+1.18}_{-1.14}$ in the 12--22 keV band.
Compiling this with the soft X-ray spectrum simultaneously observed with the X-ray imaging spectrometer (XIS) onboard Suzaku, 
we  find that  the wide-band X-ray spectrum in the 2--22 keV band is reproduced with
a single power-law or concave broken power-law model, 
which are statistically consistent with each other.
Whichever the model of a single or broken power-law is appropriate, 
clearly the spectrum has no rolloff structure.
 Applying this result to the method  introduced in \citet{yama2014}, 
we find that one-zone synchrotron model with electron spectrum having a power-law plus exponential cutoff
may not  be applicable to Vela Jr.
\end{abstract}

\section{Introduction}


Several supernova remnants (SNRs) are known as sites of acceleration  of cosmic-ray particles with the energy of up to TeV. 
The first observational evidence is discovered by \citet{koya1995} in the X-ray band.
They found X-rays from shell regions of SN 1006 with ASCA, 
and revealed that the X-rays are produced via synchrotron radiation by TeV electrons.

We have not yet succeeded observationally in determining the maximum energy and 
acceleration efficiency of protons and even electrons, which
are the key to understand the acceleration site of Galactic cosmic rays.
In the context of diffusive shock acceleration, particles with a higher energy  require a longer acceleration time,
while they have a shorter cooling and escape time  (e.g. \cite{drur2011}, \cite{laga1983}).
 These timescales are functions of a diffusion coefficient of accelerated particles around the shock 
(\cite{bell1987} and \cite{ohir2010}).
Curvature in X-ray synchrotron spectrum  depends on the energy spectrum of highest-energy electrons there. 
Measuring the spectral shape (i.e., slope or photon index, rolloff frequency and spectral curvature)
 expected to be observed in the X-ray synchrotron emission 
is crucial to understand the environment and mechanism of cosmic ray acceleration (\cite{yama2014}, \cite{yama2015}).

So far wide-band X-ray spectra 
 have been extensively studied in several SNRs; e.g., RX J1713.7$-$3946 (\cite{taka2008}, \cite{tana2008}), 
SN 1006 \citep{bamb2008}, Cassiopeia A \citep{maed2009}, and G1.9$+$0.3 \citep{zogl2015}.
Rolloff frequencies are reported from a number of SNRs.
\citet{bamb2005a} reported the rolloff energies distributing $0.07 - 11.6$ keV, from Cas~A, Kepler SNR, Tycho SNR and RCW~86.
Utilizing hard X-ray spectra obtained with NuSTAR, the rolloff energies of $> 2$~keV and $1.27\pm0.07$ keV were observed
from Tycho SNR \citep{lope2015} and G~1.9$+$0.3 \citep{zogl2015}, respectively.
TeV gamma-rays have been also observed from a number of SNRs, 
which are thought to be produced via inverse Compton scattering of TeV electrons or 
via $\pi^0$ decay process of high-energy protons (e.g., \cite{ahar2004}).

The young SNR Vela Jr.\ (RX J0852$-$4622) was discovered by ROSAT in 1998 \citep{asch1998}. It is one of  TeV-emitting SNRs, 
 and has a diameter of 2$^{\circ}$. 
It overlaps with the south-east portion of much larger and known Vela SNR, 
and  is situated near the pulsar and pulsar wind nebulae (PWN) of Vela SNR.
Vela Jr.\ has a synchrotron X-ray shell (\cite{slan2001}, \cite{bamb2005b}), which  is observed also in radio \citep{comb1999} and gamma-rays \citep{ahar2005}. 
\citet{fuku2013} reported a spatial correlation  
between the radio emission from molecular clouds around Vela Jr.\ and TeV gamma-rays.
Magnetic field strength at the shock surface of Vela Jr.\ has been estimated by some authors.
While \citet{bamb2005b} estimated the field strength, 
$B \sim 5 \times 10^{2} \mu$G from the thickness of X-ray thin shell taken by Chandra,
\citet{kish2013} derived $B \sim 5$--$20\ \mu$G, by comparing brightness distribution profile at the shell in 2--10 keV 
with expected radial profile by \citet{petr2011}. 
\citet{lee2013} constructed
1D spherically symmetric model of non-linear diffusive shock acceleration,
and obtained $B \sim 4.8\ \mu$G from observation results, assuming synchron X-rays and 
cosmic microwave background photon upscattered to TeV gamma-rays.
The observed spectral slope is so steep that we naturally expect that the synchrotron rolloff energy is 
below 1 keV, and the spectrum becomes much softer for higher energy bands.
However, the synchrotron  emission above 10 keV has not been observed for Vela Jr., 
and thus the spectral shape has not yet been determined.
In order to determine the photon index and to examine the expected rolloff structure, 
wideband X-ray observation covering above 10 keV is important.
X-ray astronomy satellite Suzaku \citep{mits2007} covers the energy range of 0.2--600 keV   with four units of 
the X-ray Imaging Spectrometers (XISs: \cite{koya2007}) and the Hard X-ray Detector (HXD: \cite{taka2007}).
Since both instruments have very low-background capabilities (\cite{tawa2008}; \cite{fuka2009}),
which enable us to perform high sensitivity surveys of X-rays, the Suzaku is most suitable for diffuse objects.
We  report  the first result of spectral analysis  of wideband observation of Vela Jr.\ with Suzaku.  
We describe Suzaku observations and data reduction of Vela Jr.\ in section 2, 
summarize results of analysis in section 3, and present a discussion in section 4.


\section{Observation and Data Reduction}

We preformed 40 mapping observations  of  Vela Jr.\ and its close vicinity
 in 2005 December, 2007 July, and 2008 July. 
Table \ref{tab:obsno} shows the observation details.
Hereafter, we refer to each observation position in an abbreviated style by clipping each object name.
For example, we call Vela Jr P1 as P1, and RXJ\_0852$-$4622\_NW as NW. 

One of the two types of instruments of Suzaku is a set of the XISs, each of which is installed
 on the focal plane of an individual set of the X-ray telescopes 
(XRTs: Serlemitsos et al. 2006\footnote{http://www.astro.isas.ac.jp/suzaku/doc/suzakumemo/suzakumemo-2006-34.pdf}).
The XIS covers a field of view (FOV) of $17'.8 \times 17'.8$ with the angular resolution of $2'$ in half power diameter 
in the energy range of 0.2--12 keV.
The XIS 0, 2, and 3 are front side illuminated (FI) CCDs, whereas XIS 1 is a back-side illuminated (BI) CCD.
FI  CCDs are more sensitive in the energy band above 5 keV because the depletion layers are thicker than  that of a BI CCD. 
We use only XIS 0, 1, and 3 data because XIS 2  has not been operated since 2007.
Spaced-row charge injections (\cite{naka2008}, \cite{uchi2009}) were carried out in all the observations except NW and NW\_offset.
We use the software package  {\tt Heasoft 6.12}  with {\tt CALDB 2009-08-04} for the analysis
and {\sc XSPEC} v12.7.1 for the spectral analysis\footnote{https://heasarc.gsfc.nasa.gov/xanadu/xspec/}.
Each observation data is reprocessed by {\tt aepipeline 1.0.1}.
We extract events based on the following criteria: 
elevation angle from night earth $> 5^{\circ}$,
elevation angle from day earth $> 20^{\circ}$.
We remove calibration source regions at the corner of the FOV.

The HXD is a well-type phoswitch counter, whose main detection-part consists of silicon detectors and GSO crystal scintillators.
The silicon PIN type semiconductor detector covers hard X-ray band of 10--70 keV, 
and has FOV of $34' \times 34'$.
The GSO crystal scintillators cover the band of 40--600 keV.
We do not present the result of GSO  because no significant detection is made.
Since the well-type active shield provides low-background environment for the PIN detector, 
it is one of the ideal detectors to observe low surface-brightness objects like SNRs.
We extract events  which have an elevation angle of $> 5^{\circ}$  and geomagnetic cut-off rigidity of $> 6$ GV.

\begin{longtable}{llllll}
  \caption{The basic parameters of the Suzaku observations. 
    The exposures  are the ones after processing and 
     of the XIS, unless otherwise noted. }
  \label{tab:obsno}
  \hline              
  Name & ObsID & Date (YYYY-MM-DD) & RA (deg) & Dec (deg) & Exposure (ks) \\ 
  \hline
\endfirsthead
  \hline
  Name & ObsID & Date (YYYY-MM-DD) & RA (deg) & Dec (deg) & Exposure (ks) \\ 
  \hline
\endhead
  \hline
\endfoot
  \hline
\endlastfoot
      VELA JR P1  & 502023010 & 2007-07-04 & 131.98 & $-$45.806 & 10.7 \\
      VELA JR P2  & 502024010 & 2007-07-04 & 132.17 & $-$45.775 & 8.26 \\
      VELA JR P3  & 502025010 & 2007-07-04 & 132.12 & $-$45.604 & 6.70 \\
      VELA JR P4  & 502026010 & 2007-07-05 & 132.52 & $-$45.545 & 10.3 \\
      VELA JR P5  & 502027010 & 2007-07-05 & 132.91 & $-$45.488 & 10.7 \\
      VELA JR P6  & 502028010 & 2007-07-05 & 133.33 & $-$45.485 & 7.15 \\
      VELA JR P7  & 502029010 & 2007-07-05 & 133.78 & $-$45.583 & 11.8 \\
      VELA JR P8  & 502030010 & 2007-07-06 & 133.86 & $-$45.861 & 13.2 \\
      VELA JR P9  & 502031010 & 2007-07-06 & 133.42 & $-$45.763 & 8.64 \\
      VELA JR P10 & 502032010 & 2007-07-06 & 133.00 & $-$45.766 & 10.2 \\
      VELA JR P11 & 502033010 & 2007-07-07 & 132.60 & $-$45.826 & 11.3 \\
      VELA JR P12 & 502034010 & 2007-07-08 & 132.25 & $-$46.051 & 9.75 \\
      VELA JR P13 & 502035010 & 2007-07-09 & 131.85 & $-$46.106 & 9.41 \\
      VELA JR P14 & 502036010 & 2007-07-09 & 131.93 & $-$46.386 & 10.7 \\
      VELA JR P15 & 502037010 & 2007-07-10 & 132.33 & $-$46.329 & 8.88 \\
      VELA JR P16 & 502038010 & 2007-07-10 & 132.68 & $-$46.105 & 15.1 \\
      VELA JR P17 & 502039010 & 2007-07-10 & 133.09 & $-$46.046 & 7.83 \\
      VELA JR P18 & 502040010 & 2007-07-10 & 133.51 & $-$46.042 & 12.8 \\
      VELA JR P19 & 503031010 & 2008-07-03 & 133.98 & $-$46.148 & 17.7 \\
      VELA JR P20 & 503032010 & 2008-07-04 & 133.62 & $-$46.327 & 13.4 \\
      VELA JR P21 & 503033010 & 2008-07-04 & 133.20 & $-$46.330 & 11.8 \\
      VELA JR P22 & 503034010 & 2008-07-05 & 132.44 & $-$46.613 & 14.7 \\
      VELA JR P23 & 503035010 & 2008-07-05 & 132.79 & $-$46.395 & 10.7 \\
      VELA JR P24 & 503036010 & 2008-07-05 & 132.03 & $-$46.673 & 12.4 \\
      VELA JR P25 & 503037010 & 2008-07-06 & 132.52 & $-$46.892 & 11.4 \\
      VELA JR P26 & 503038010 & 2008-07-06 & 132.88 & $-$46.669 & 10.7 \\
      VELA JR P27 & 503039010 & 2008-07-06 & 133.28 & $-$46.606 & 11.0 \\
      VELA JR P28 & 503040010 & 2008-07-07 & 133.71 & $-$46.604 & 10.9 \\
      VELA JR P29 & 503041010 & 2008-07-07 & 134.07 & $-$46.430 & 8.04 \\
      VELA JR P30 & 503042010 & 2008-07-07 & 134.17 & $-$46.704 & 10.1 \\
      VELA JR P31 & 503043010 & 2008-07-08 & 133.80 & $-$46.885 & 10.6 \\
      VELA JR P32 & 503044010 & 2008-07-08 & 133.37 & $-$46.887 & 7.64 \\
      VELA JR P33 & 503045010 & 2008-07-08 & 133.47 & $-$47.162 & 11.0 \\
      VELA JR P34 & 503046010 & 2008-07-09 & 132.11 & $-$46.950 & 10.6 \\
      VELA JR P35 & 503047010 & 2008-07-09 & 132.97 & $-$46.947 & 8.65 \\
      VELA JR P36 & 503048010 & 2008-07-09 & 133.06 & $-$47.224 & 9.36 \\
      VELA JR P37 & 503049010 & 2008-07-09 & 132.61 & $-$47.170 & 12.3 \\
      VELA JR P38 & 503050010 & 2008-07-10 & 132.20 & $-$47.231 & 10.4 \\
      RXJ\_0852$-$4622\_NW & 500010010 & 2005-12-19 & 132.29 & $-$45.616 & 161 (XIS) / 215 (PIN) \\
      RXJ\_0852$-$4622\_NW\_offset & 500010020 & 2005-12-23 & 135.13 & $-$47.910 & 54.5 \\
      \hline
\end{longtable}

\section{Spectral analysis}\label{sec:ana}

\subsection{XIS}\label{sec:xisspec}

\begin{figure}
  \begin{center}
    \includegraphics[width=8cm]{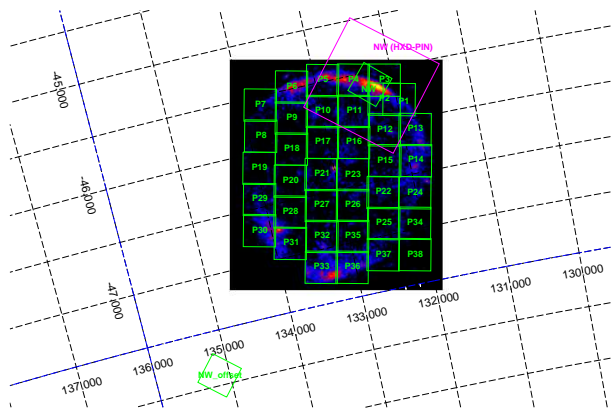}
  \end{center}
  \caption{The mosaiced XIS image of Vela Jr.\ in 2--5 keV band.  
    The exposure is corrected, but the CXB is not subtracted.
    The regions of the calibration sources are removed.
    The green and magenta boxes show the FOVs of the XIS of all the observations and  of the PIN during NW observation, respectively.
    The coordinates are in the J2000 equatorial system.
  }\label{fig:xisimg}
\end{figure}

We begin with the analysis of the XIS data.
Figure \ref{fig:xisimg} shows the mosaiced XIS image of Vela Jr.\ in 2--5 keV,
which is created with {\tt ximage}, combining the exposure and vignetting-effect corrected image of each observation.
Shell structures of Vela Jr.\ are clearly visible.
We focus on the NW shell which is detected in TeV gamma-rays  \citep{kata2005} and 
reportedly shows bright synchrotron X-ray filaments \citep{bamb2005b}.
In order to  match the data with those of the HXD-PIN, 
we  selected the region inside the  FOV of the PIN NW observation 
which is a square sky region with apexes of
(133.37, $-$45.69), (132.17, $-$46.45), (131.09, $-$45.61) and (132.29, $-$44.86) 
in equatorial coordinates.  
Consequently, the entire regions of NW, P1, P2, P3, P4, P11 and P12, 
and parts of P5, P6, P9, P10, P13, P14, P15, P16 and P17 are included.
 The spectra are summed with {\tt mathpha} in  units of counts without exposure weighting.
The errors are propagated as Poisson errors.
Response files are created with {\tt marfrmf} from redistribution matrix files (rmfs) made by {\tt xisrmfgen} 
and ancillary response files (arfs) made by {\tt xissimarfgen} \citep{ishi2007}, based on the XIS 2--5 keV image file within the FOV of the PIN NW observation.
We add all the XIS responses of each observation with the weight of each exposure.

\begin{figure}
  \begin{center}
    \includegraphics[width=8cm]{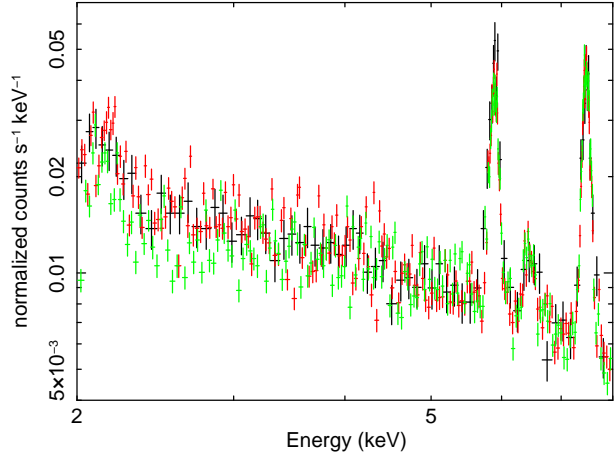}
  \end{center}
  \caption{Background spectrum of the XIS 0 in 2.0--8.0 keV band.
    The black, red, and green lines denote the NW\_offset, the NXB-replaced background (A), and the simulated NXB plus modeled CXB  \citep{kush2002} background (B), respectively.
  }\label{fig:xisbgspec}
\end{figure}

Since the object extends toward the outside of XIS FOV, 
we should evaluate background spectra carefully.
We here employed the XIS spectra from the observation 
RXJ\_0852$-$4622\_NW\_offset (table~\ref{tab:obsno}: NW\_offset) as the background spectra.
However, we must consider the difference in NXB by the difference of the observation days.
The NW\_offset spectra were obtained in winter of 2005, 
while the source spectra are the sum of observations in winter of 2005 and summer of 2007.
In order to evaluate the possible temporal variation of NXB, we made background spectra (A) by replacing NXB component.
The NXB component of each occation is reproduced by {\tt xisnxbgen}
which makes use of the NXB database of the night-earth observations by \citet{tawa2008}
extracting the data from 4 months before to 16 months after an observation.
In addition, to examine possible contaminate source in NXB\_offset FOV, 
we compare the background spectra with the NXB plus the CXB estimated by \cite{kush2002} (B).
Above backgrounds are summarized in figure \ref{fig:xisbgspec} and we confirmed
the background spectra are consistently reproduced within the errors.
Then, we safely adopt the NW\_offset background 
ignoring the energy range of 5.9 $\pm$ 0.2 keV also
to eliminate scattered $^{55}$Fe line from the decaying calibration source during the observations.

\begin{figure}
  \begin{center}
    \includegraphics[width=8cm]{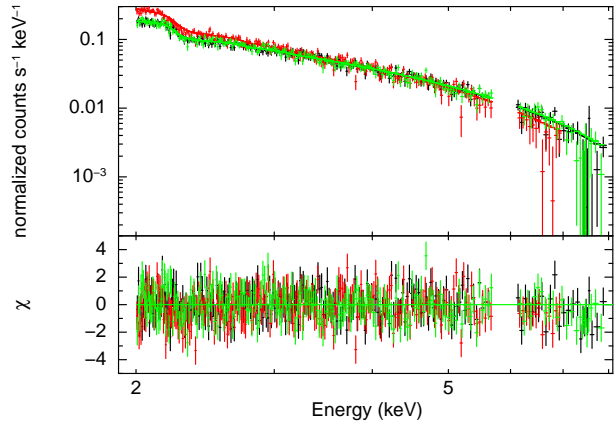}
  \end{center}
  \caption{The upper panel shows the XIS spectra fitted with a single power-law model in the 2.0--8.0 keV band.
    The lower panel shows the $\chi$-residuals between the data and best-fit model.
    The black, red, and green lines  denote the XIS 0, 1, and 3 data, respectively.
  }\label{fig:xisspec}
\end{figure}

We evaluate background-subtracted XIS spectra
by fitting with a power-law function with the Galactic absorption: {\tt phabs*powerlaw}.
The absorption column density is fixed to be $6.7 \times 10^{21}$ cm$^{-2}$,
which is the best-fit value  derived by \citet{hira2009}, with the metal abundance adopted from \citet{ande1989}.
We ignore the energy range of below 2 keV to avoid thermal contamination from Vela SNR \citep{hira2009}.
Figure \ref{fig:xisspec} shows the XIS spectra, and Table \ref{tab:xispinbestfitpara} shows the best-fit parameters.

\begin{table}[h]
\caption{
The best-fit parameters for a power-law model for the XIS spectra and the PIN spectrum.
}\label{tab:xispinbestfitpara}
\begin{center}
\begin{tabular}{lll} \hline
Parameter                   & XIS\footnotemark[$*$] & HXD-PIN\footnotemark[$\dagger$]  \\   \hline
Photon index                & $2.93 \pm 0.02$       & $3.15^{+1.18}_{-1.14}$  \\
flux [erg cm$^{-2}$ s$^{-1}$] & $4.43 \pm 0.03 \times 10^{-11}$\footnotemark[$\ddagger$] & $8.26 \pm 1.44 \times 10^{-12}$\footnotemark[$\S$] \\
$\chi^{2}$/d.o.f.            & $732.16 / 658$  & $0.87/4$            \\   \hline
  \multicolumn{3}{@{}l@{}}{\hbox to 0pt{\parbox{170mm}{\footnotesize
        Notes. Errors are for a single parameter  of interest in 90\% confidence.
    }\hss}}
\end{tabular}
\end{center}
\begin{tabnote}
\hangindent6pt\noindent
\hbox to6pt{\footnotemark[$*$]\hss}\unskip%
The absorbing column density is set to be $6.7 \times 10^{21} \rm cm^{-2}$,  referring to \citet{hira2009}. \\
\hbox to6pt{\footnotemark[$\dagger$]\hss}\unskip%
The  model contains the fixed GRXE  component  and the normalization is  corrected (see text). \\
\hbox to6pt{\footnotemark[$\ddagger$]\hss}\unskip%
The flux at the range of 2--10 keV. \\
\hbox to6pt{\footnotemark[$\S$]\hss}\unskip%
The flux at the range of 12--22 keV.
\end{tabnote}
\end{table}

\subsection{HXD-PIN}\label{sec:pinspec}

To identify pointings which show significant hard X-ray  signals with HXD-PIN, we compare the  background-subtracted  count-rates of each observation 
with the systematic error of  the corresponding simulated background spectrum.
We  employ  a modeled CXB by  \citet{bold1987} and simulated NXB spectrum  with {\tt hxdpinxbpi}, the latter of which
 is calculated on the basis of observed NXB spectrum during earth occultation.
The expected uncertainty of the NXB model is reported as 3\% in 10--60 keV by \citet{fuka2009}.
The PIN detection significance is determined  with the NXB reproducibility.
Since hard X-ray sources RCW 38 and IGR J09026--4812 contaminate the NW\_offset observation,
those offset observations  are not to  used to  estimate the NXB and CXB backgrounds for this HXD-PIN analysis.
We ignore the energy band below 12 keV to avoid thermal noise.
Consequently, we obtain 6 observations that exceed 1 $\sigma$ significance level of NXB uncertainty in the 10--60 keV band: 
P1, P2, P3, P5, P13, and NW,
all of which are located in the north-west of Vela Jr.

To confirm the detections, we re-evaluate uncertainty of the NXB model for each observation by comparing 
the count rate of CXB-subtracted signal 
with that of each simulated NXB  in the 10--60 keV  band.
 For the NXB, we derive the count rate  during earth occultations during each observation.
 All but  the NW contained some periods of earth occultations.
 For the NW observation,
we used the earth occultation data obtained in the observations conducted  immediately before and 26 hours after the NW observation:
E0102$-$72 (ObsID 100044010) and NGC 4388 (ObsID 800017010). 
Comparing those with the count rate of the model,  we  estimate the systematic uncertainties of the NXB model  to be $\sim$ 7\%, 6\%, 0.2\%, 0.5\%, 9\%, and 1\% 
for positions P1, P2, P3, P5, P13, and NW, respectively, in 10--60 keV.
P1, P2, and P13  have larger uncertainty than the nominal value reported in \citet{fuka2009} 
because  the exposures of earth occultation were short.
 A significant emission to 21.9 keV from the position NW  is detected  at 3 $\sigma$ confidence level,
 whereas those  from the other observations  are detected at only 0.5--1.5 $\sigma$ level.

\begin{figure}
  \begin{center}
    \includegraphics[width=8cm]{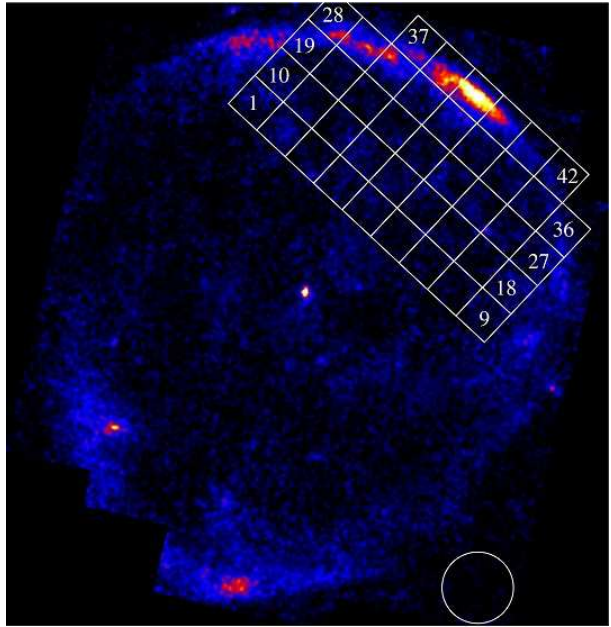}
  \end{center}
  \caption{Forty-two white boxes  used to estimate the arf, in which the significant X-ray emission was detected with the XIS (see text).
    A white circle is the region  for estimating the background.
    The image is the same as Figure \ref{fig:xisimg}.
  }\label{fig:pinarf}
\end{figure}

In order to examine the PIN spectrum of diffuse objects, 
we need to calculate the effective areas for which the angular response is convolved 
(`arf' in the {\sc XSPEC}), 
 based on the source brightness distribution within the FOV.
We assume that the spatial distribution in the hard X-ray band 
with PIN is the same as the CXB-subtracted XIS 2--5 keV images 
(section~\ref{sec:xisspec}).
The proper response of the PIN detectors has a pyramidal shape.
However, for simplicity, we divide the spatial distribution within the PIN FOV by $9 \times 9$ grids  and 
make 81 arfs in total with {\tt hxdarfgen}, assuming that a point source centered in each section is responsible for the entire flux from the section  in making an arf in each section.
Then we sum up these arfs with {\tt addarf} with weights calculated from the XIS image. 
Practically, this procedure is to approximate the original pyramidal angular response by $9 \times 9$ prisms. 
We calculate  the ratio of the geometric integrations of the 81 prisms to the pyramid  to be  0.8.  Thus, the flux based on this arf should come out at the value $1 / 0.8 = 1.25$ times larger than the real one.
\newline \indent We validate this ratio of the normalizations with another pair of data sets: (A)  the effective area derived with this method  for the $9 \times 9$ section with a uniform weight, (B)
 that calculated from the numerically-simulated flat-sky response file, which is delivered by the HXD team. 
 We confirm that  the ratio of the former (A) to the latter (B) is $\sim$0.8 and so is consistent with  the value calculated  above. \newline \indent
 Now that the method is validated, we make the arf for our HXD-PIN spectrum with this method, assuming the HXD-PIN spatial distribution of Vela Jr.\ to be the same as  the CXB-subtracted brightness distribution  observed with the XIS.
We  find that 39 out of the 81 sections inside Vela Jr.\ show no significant XIS signal and hence  give them the weight of zero in calculating the arf for the HXD-PIN spectrum.
Figure \ref{fig:pinarf}  displays the employed sections, and Table \ref{tab:pinarf}  lists  the weights. All the flux and flux values presented hereafter are corrected for the above-mentioned factor.

Using the arf for Vela Jr.\ NW with the officially delivered  responses for the flat sky and a point source, 
we  examine the observed hard X-ray spectrum in detail.
First,  we  evaluate possible contamination 
of the galactic ridge X-ray emission (GRXE: \cite{kriv2007}), given the fact that Vela Jr.\ is on the Galactic plane.
We use the NXB-subtracted XIS spectrum of the offset observation as the background template for the HXD data of Vela Jr.
We fit  it with, in addition to the fixed CXB model \citep{kush2002},  the GRXE model, for which
we employ a photo-absorbed two temperature thermal plasma emission model.
Assumed absorption column density $N_{\rm H} = 4.0 \times 10^{22} {\rm cm}^{-2}$ and the 
two temperature thermal plasma model ({\tt apec} in {\sc XSPEC}) with $kT = 1.66$ keV and $15.1$ keV (Table 4 of \cite{yuas2012}).
Considering the effective solid angle of PIN and XIS, we then estimate the GRXE component flux for the PIN spectrum (12 -- 22 keV) and XIS spectrum (2 -- 10 keV) to be 
$2.00 \times 10^{-13}$ and $7.01 \times 10^{-13}$ ${\rm erg\ cm}^{-2} {\rm s}^{-1}$, respectively.
This estimated GRXE model is included in all the following model-fittings of the HXD-PIN spectrum.
Note that 
the flux of GRXE in the XIS range is 2\% of  that from NW. 
Therefore, the effect of the GRXE to the XIS spectrum is negligible. 

 Second, we check the possible contamination from  nearby hard X-ray sources.
According to the INTEGRAL catalog\footnote{http://www.isdc.unige.ch/integral/science/catalogue}, 
there  was no contaminating point-source in the 18.3--59.9 keV range in the FOV of the HXD-PIN.
The brightest diffuse source in the PIN FOV is Vela PWN below 10 keV \citep{kats2011}.
\citet{mori2014} studied the spectrum of Vela PWN with the XIS
 observation of VELA PWN E7 (ObsID 506050010), 
 and reported that the surface brightness is  $\sim 0.3$ times  CXB level and that the photon index is $\sim 3.3$.  This flux corresponds to
5\% of that of the NW in the 2--10 keV band, and hence 
our XIS spectrum is not significantly affected with the Vela PWN component within statistics.
Extrapolating this spectrum to the energy band of the HXD-PIN, 
we  find that the flux of Vela PWN is $\leq$ 2\% of that of the NW in the 12--22 keV energy range, and therefore the contribution from the Vela PWN component is negligible for the PIN data.
Therefore, we conclude that the detected signals are fully originated from Vela Jr.\ NW.

\begin{figure}
  \begin{center}
    \includegraphics[width=8cm]{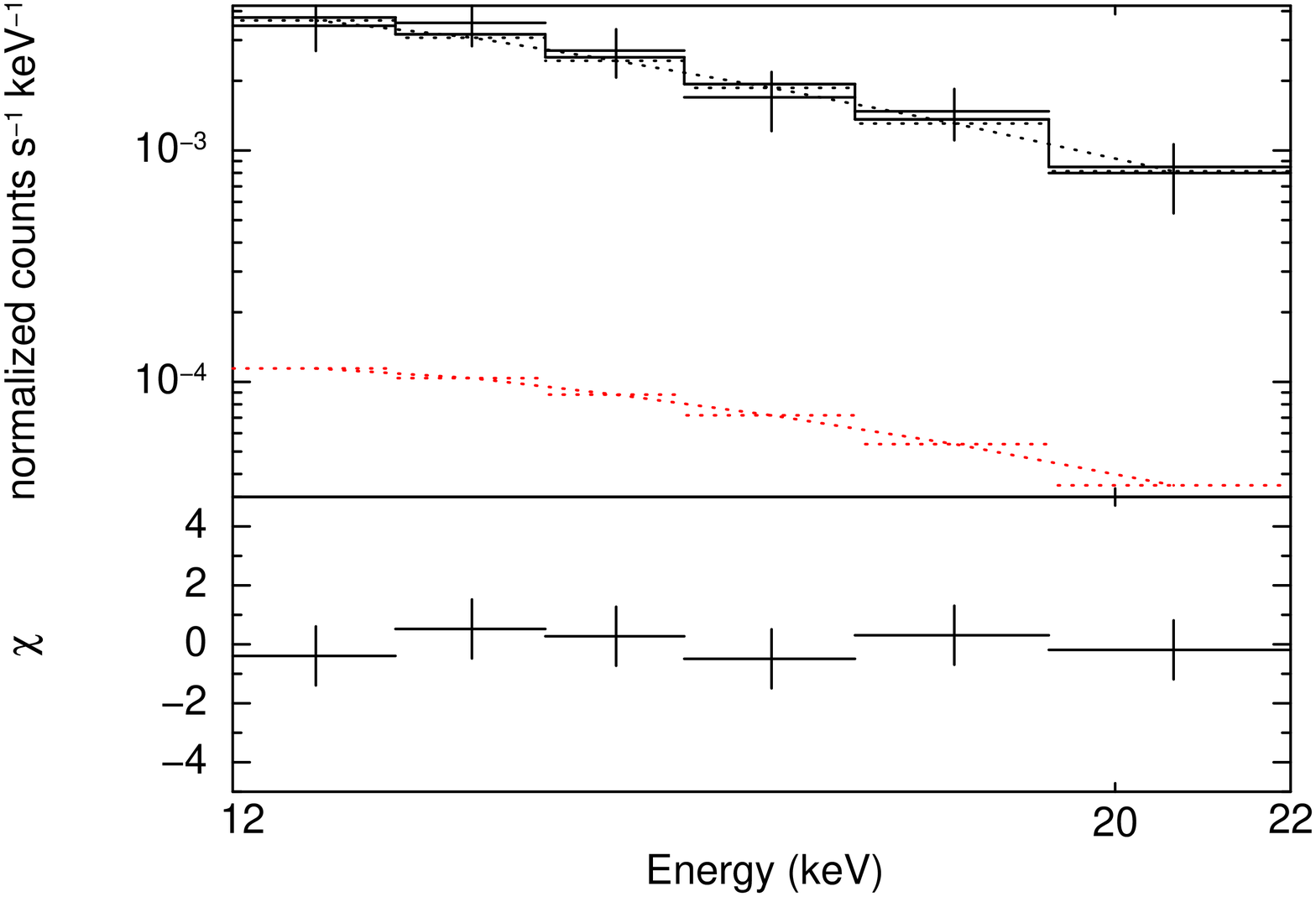}
  \end{center}
  \caption{The black points in the upper panel show the HXD-PIN spectrum fitted with 
    single power-law model in 12.0--22.0 keV band.
    The error bars include also the systematic error of the NXB model.  
    The red dotted line shows  the contribution from the modeled GRXE. 
    The lower panel shows the $\chi$ residuals between the data and best-fit model.
  }\label{fig:pinspec}
\end{figure}

Finally we fit the background-subtracted HXD-PIN spectrum up to 22 keV with a single power-law model.
Figure \ref{fig:pinspec} shows the  spectrum 
with the best-fit model. 
The  systematic error of the NXB model of 1\%, which is derived  based on the earth occultation data, 
 is included in errors of the source spectrum.
The spectrum  is well reproduced by a single power-law model with  $\chi^{2}$/d.o.f. $\sim 0.22$.
The  best-fit photon index is $3.15^{+1.18}_{-1.14}$
and the flux  is $(8.26 \pm 1.44) \times 10^{-12}$ erg cm$^{-2}$ s$^{-1}$ in 12--22 keV, 
as listed in table~\ref{tab:xispinbestfitpara}.  
Here the  errors due to the model-fitting error of the GRXE are 0.006\%  for the photon index and 0.02\%  for normalization.

\begin{table}
  \caption{The weights for making arfs of the PIN. The pointing  IDs are shown in figure \ref{fig:pinarf}.
  }\label{tab:pinarf}
  \begin{center}
    \begin{tabular}{rlrl}
      \hline
       Pointing ID & weight &  Pointing ID & weight \\ 
      \hline
      1 & 0.018 & 22 & 0.012 \\
      2 & 0.020 & 23 & 0.013 \\
      3 & 0.012 & 24 & 0.017 \\
      4 & 0.014 & 25 & 0.016 \\
      5 & 0.013 & 26 & 0.016 \\
      6 & 0.009 & 27 & 0.021 \\
      7 & 0.009 & 28 & 0.033 \\
      8 & 0.009 & 29 & 0.068 \\
      9 & 0.015 & 30 & 0.053 \\
      10 & 0.019 & 31 & 0.028 \\
      11 & 0.013 & 32 & 0.032 \\
      12 & 0.014 & 33 & 0.016 \\
      13 & 0.017 & 34 & 0.016 \\
      14 & 0.015 & 35 & 0.018 \\
      15 & 0.013 & 36 & 0.021 \\
      16 & 0.013 & 37 & 0.034 \\
      17 & 0.009 & 38 & 0.076 \\
      18 & 0.022 & 39 & 0.119 \\
      19 & 0.034 & 40 & 0.051 \\
      20 & 0.013 & 41 & 0.022 \\
      21 & 0.010 & 42 & 0.008 \\
      \hline
    \end{tabular}
  \end{center}
\end{table}

\subsection{Wide-band spectrum}\label{sec:widespec}


In sections \ref{sec:xisspec} and \ref{sec:pinspec}, we have derived the spectral parameters independently with the XIS and PIN.  
We then fit the XIS and PIN spectra simultaneously to give further constraints, making use of the best available statistics.

\begin{figure}
  \begin{center}
    \begin{tabular}{cc}
      \resizebox{80mm}{!}{\includegraphics{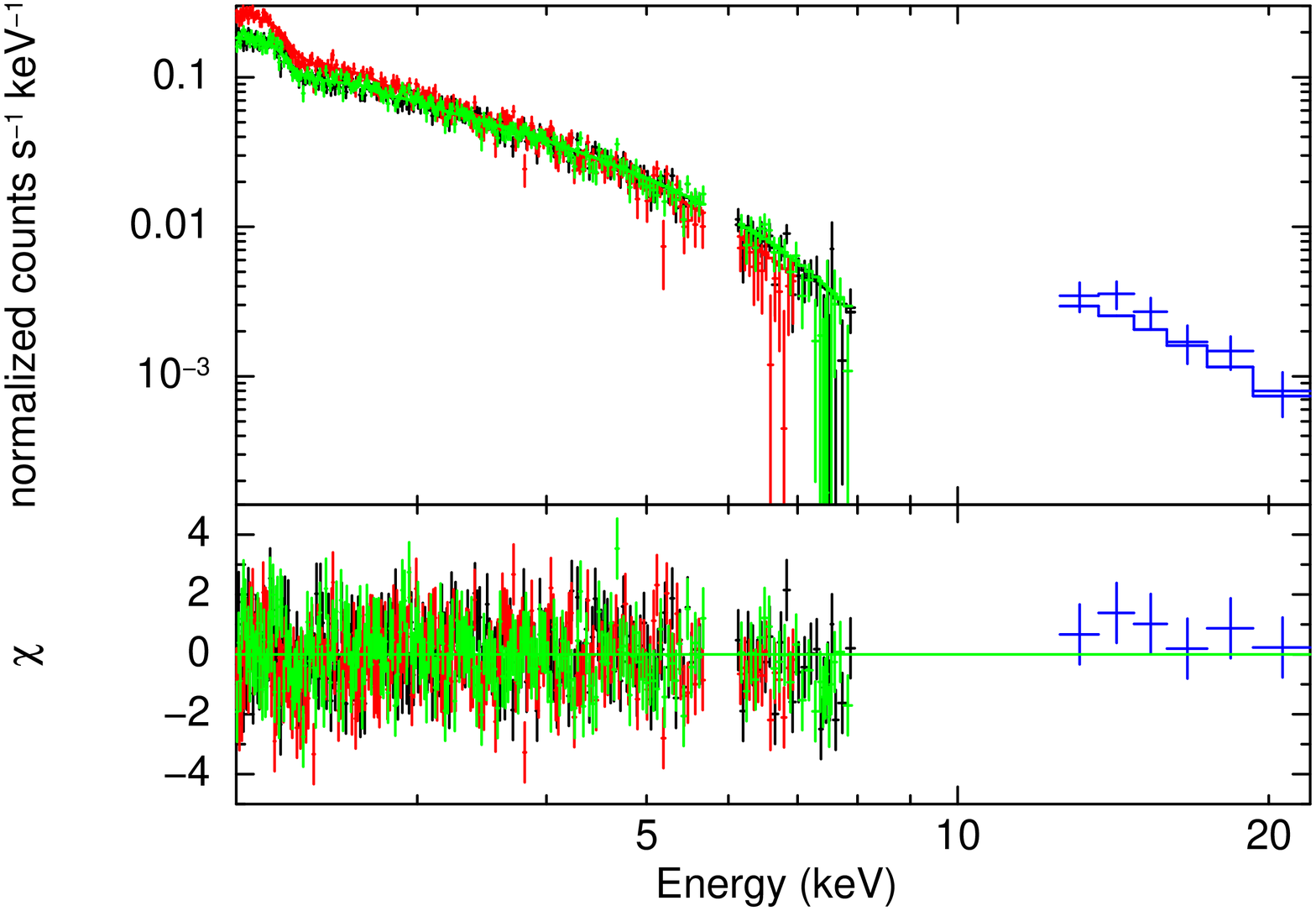}} &
      \resizebox{80mm}{!}{\includegraphics{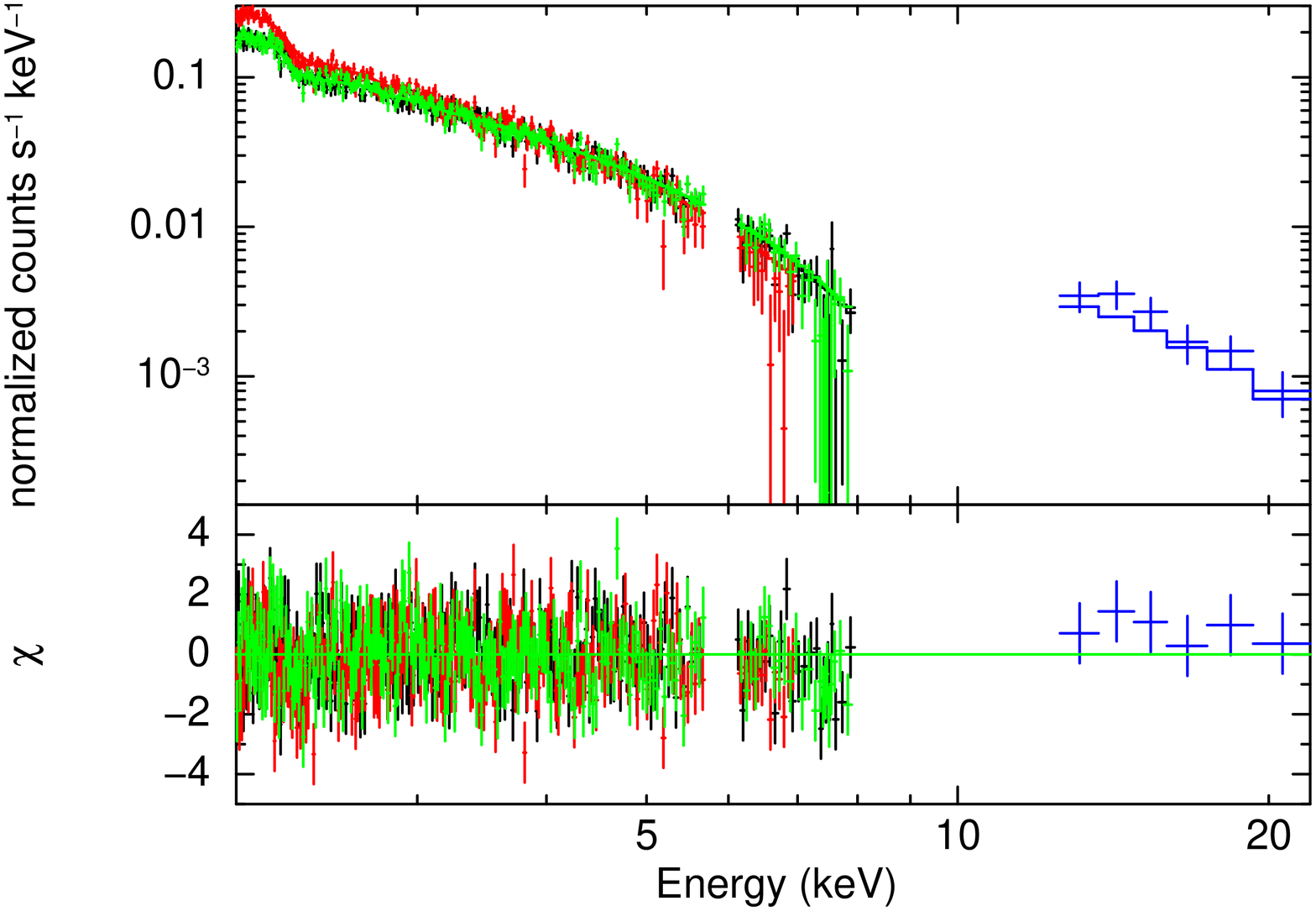}} \\
      \resizebox{80mm}{!}{\includegraphics{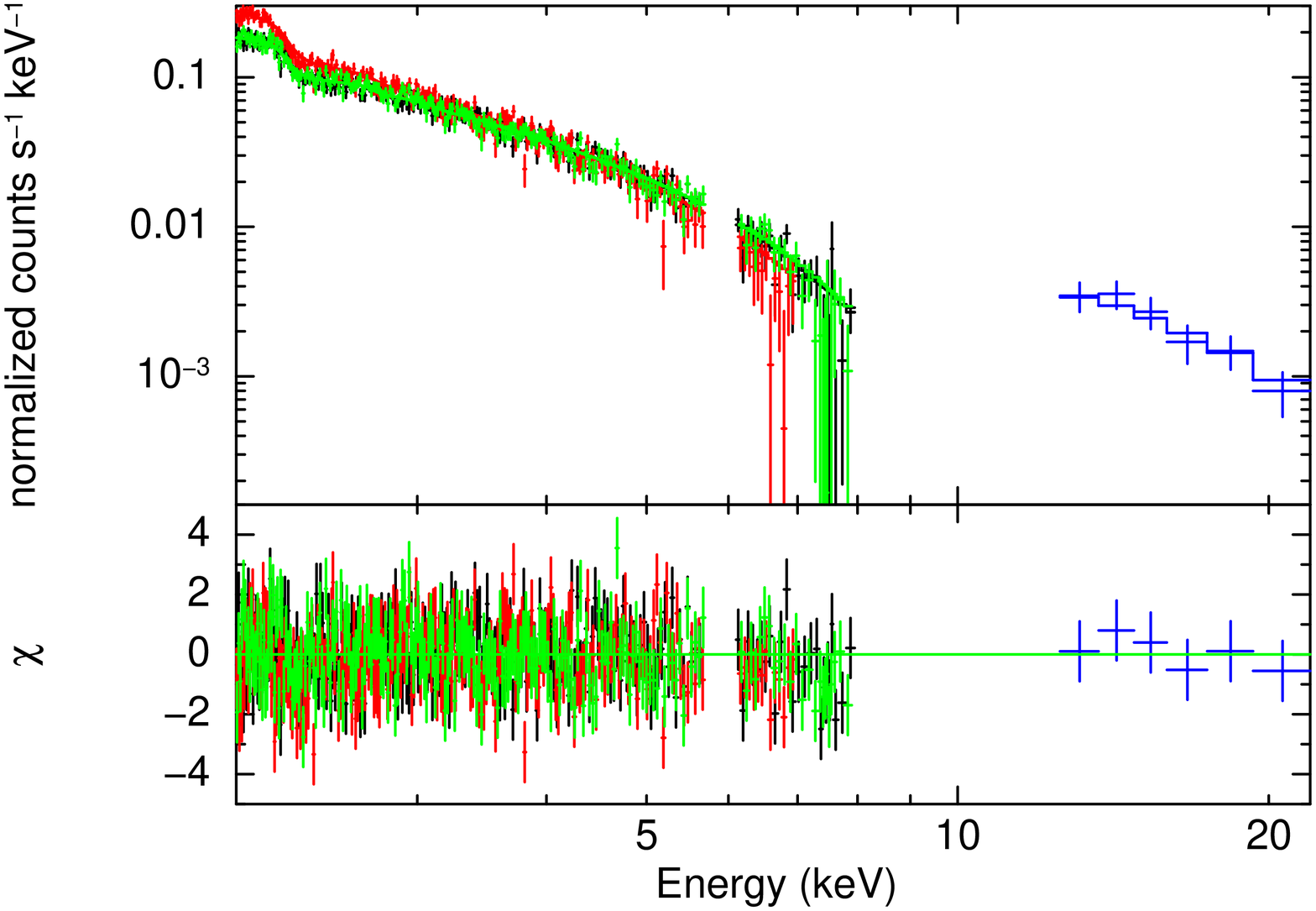}} &
      \resizebox{80mm}{!}{\includegraphics{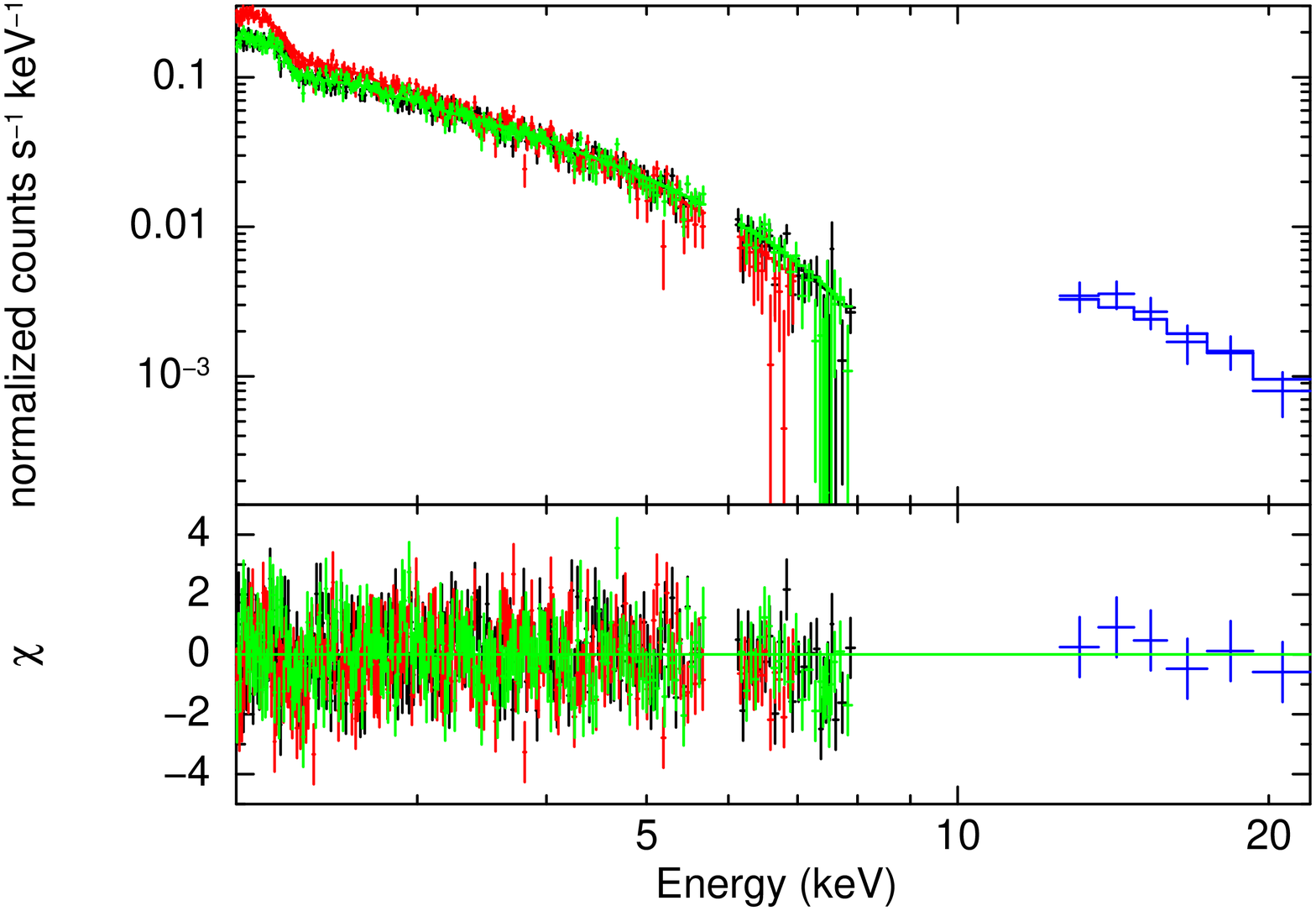}} \\
    \end{tabular}
  \end{center}
  \caption{Upper left: The upper panel shows the wide band spectrum fitted with (i) single power-law model in 2.0--22.0 keV band.
    The systematic error of NXB model is included.
    The cross normalization is fixed at 1.13, which is the value  for a point source.
    The lower panel shows the residuals between the data and model.
    Upper right: Same as the upper left except for using (ii) cutoff power-law model.
    Lower left: Same as the upper left except for using (iii) broken power-law.
    Lower right: Same as the upper left except for using (iv) 10 keV broken power-law.
}\label{fig:widespec}
\end{figure}

First, we  apply a single power-law model, referred to as model (i) in Table \ref{tab:widebestfitpara},
 linking the photon index  for the XIS and PIN spectra.
Ishida et al. (2007)\footnote{http://www.astro.isas.ac.jp/suzaku/doc/suzakumemo/suzakumemo-2007-11.pdf} reported that 
the normalization factors differed between the XIS and PIN spectra for the point-like source Crab, 
which has a power-law spectrum, and that their ratio of the PIN to the XIS (henceforth referred to as the cross-normalization) was 1.13.   
Thus, we link the power-law normalizations of the XIS and PIN spectra with the ratio of 1/1.13 in the model-fitting.
Figure \ref{fig:widespec} shows the spectra and the best-fit  model, 
and Table \ref{tab:widebestfitpara}  lists the best-fit parameters.

The X-ray spectra of several SNRs have rolloff structures (e.g. \cite{taka2008}, \cite{tana2008}, \cite{bamb2008}, \cite{zogl2015}).
Although the above result on a single power-law model does not require any spectral bending in the 2--22 keV band, 
we further try to examine a possible spectral curvature with various models listed in Table \ref{tab:widebestfitpara}, 
setting the cross-normalization to 1.13 (see the previous sub-section).
Table \ref{tab:widebestfitpara} shows the best-fit parameters of each model.
Four panels in figure \ref{fig:widespec} show the spectrum  overlaid with the best-fit model spectra for 4 different models.
The parameters  with the cutoff power-law ({\tt cutoffpl} in {\sc XSPEC}) 
which is a power-law model with high energy exponential rolloff (ii) 
are consistent with the results  with the single power-law model (i), 
because the lower limit of rolloff energy of 131 keV is out of the range.
Fitting with the broken power-law model (iii; {\tt bknpower} in {\sc XSPEC}) implies 
the breaking energy  of $7.90 \pm 0.18$ keV with a change of spectral index of $-$0.3, 
although it is still consistent with the results of individual fittings  with the XIS and PIN, 
and with the wide-band fittings with the models (i) and (ii), within  errors.

\begin{table}[h]
\caption{
  The best-fit parameters for the XIS and PIN spectra.
}\label{tab:widebestfitpara}
\begin{center}
\begin{tabular}{llllp{25mm}} \hline
Parameters                                              & (i) single power-law & (ii) cutoff power-law & (iii) broken power-law & (iv) 10 keV broken power-law \\ \hline
Photon index (all or soft)                              & $2.92 \pm 0.01$      & $2.90 \pm 0.01$       & $2.93 \pm 0.01$        & $2.93 \pm 0.01$  \\ 
Photon index (hard)                                     & --                   & --                    & $2.66 \pm 0.03$        & $2.56_{-0.34}^{+0.42}$   \\  
rolloff energy [keV]                                    & --                   & $> 131$               & --  & --  \\ 
breaking energy [keV]                                   & --                   & --                    & $7.90 \pm 0.18$      & 10.0 (fixed)  \\ 
flux$_{10 \rm{keV}}$ [$\times 10^{-5}$ Jy]                   & $5.05 \pm 0.03$      & $4.92 \pm 0.03$       & $5.30 \pm 0.03$      & $4.99 \pm 0.03$   \\ 
$\chi^{2} /$ d.o.f.                                      & $736.51/664$         & $735.69/663$          & $733.59/662$         & $733.85/663$  \\ \hline 
  \multicolumn{5}{@{}l@{}}{\hbox to 0pt{\parbox{170mm}{\footnotesize
        Notes.
        The absorbing column density is set to be $6.7 \times 10^{21} \rm cm^{-2}$,  referring to \citet{hira2009}.
        Cross-normalization factor of HXD-PIN is fixed at 1.13, which is the value  for a point source.
        Errors are for a single parameter  of interest in 90 \% confidence.
    }\hss}}
\end{tabular}
\end{center}
\end{table}

\newpage

\section{Discussion and Summary}

In section \ref{sec:ana}, we  have shown the results of spectral analysis of Vela Jr.\ with Suzaku.
 X-rays with the energy up to 22 keV  are detected from the north-west region of Vela Jr.
The spectra in the soft and hard X-ray band are reproduced  with a power-law model with the photon indices of 
$2.93 \pm 0.02$ and $3.15^{+1.18}_{-1.14}$,  respectively.
When the spectra in both the bands  are fitted simultaneously,
a single power-law model with the photon index of $2.92 \pm 0.01$, 
or a slightly concave-shaped broken power-law model,  is accepted.

The obtained wide band steep power-law like spectrum with photon index $\sim 3$ 
implies the energy index $p \sim 5$ of synchrotron electron with the energy distribution of 
${\rm d}N/{\rm d}E \propto E^{-p}$.
This steep spectrum strongly suggests that the rolloff energy is well below the XIS band.
\citet{comb1999} reported the radio spectral index of $\alpha = 0.3 \pm 0.3$ at NW region,
and \citet{dunc2000} reported the flux density at 1 GHz of $\sim$ 50 Jy from entire region of the SNR.
In order to verify consistency between the X-ray and radio results,
the X-ray spectra with XIS and PIN are tested with {\tt srcut} model in {\sc XSPEC} \citep{reyn1998}.
The {\tt srcut} model describes the synchrotron spectrum from electrons with
an exponentially-rolloff power-law distribution in energy.
The synchrotron spectrum has a power-law form from radio band to X-ray band 
with a rolloff energy.
The {\tt srcut} model has three parameters; X-ray rolloff energy, 
spectral index at 1 GHz, and flux at 1 GHz.
We fit the X-ray spectra using spectral index $\alpha$ in radio band reported by \citet{comb1999}
and typical value of young SNRs, i.e., $\alpha = 0.3$ and $\alpha = 0.6$, respectively.
Then, when $\alpha = 0.3$, the best fit parameters of 1 GHz flux density and rolloff energy of $0.15 \pm 0.01$ Jy
and $0.15 \pm 0.01$ keV with $\chi^{2}$/d.o.f. of 1.18.
When $\alpha = 0.6$, these are $31.6^{+1.8}_{-1.9}$ Jy and $0.27 \pm 0.01$ keV with $\chi^{2}$/d.o.f. of 1.17.
The area of NW region in this paper is $\sim$ 27 \% of entire Vela Jr.,
hence the 1 GHz flux is expected to become smaller than values reported by \citet{dunc2000}
which is derived from emission of entire Vela Jr.
Actually the derived radio flux is smaller than the value of entire Vela Jr.\ reported by \citet{dunc2000}.
In addition, rolloff energy is below the XIS band, which is consistent with our X-ray analysis.
Therefore results of our X-ray analysis and in radio band are consistent each other.
Since the radio flux at NW region alone has not been reported,
additional radio observation is needed to verify this result on wideband spectrum of the NW region.

Thus the steep X-ray spectrum naturally
requires concave rolloff structure as far as we assume 
simple acceleration/synchrotron cooling mechanism. 
For example, \citet{yama2014} proposed a simple diagnostic to find possible acceleration mechanisms
from the observed spectral shape near the maximum energy, assuming (a) one-zone, (b) electron energy spectrum of an exponential cutoff power-law ($N(E) \propto E^{-p} \exp [-(E / E_{\rm max,e})^{a}]$) expressed as 
equation 1 in \citet{yama2014}, 
and (c) synchrotron radiation.
Figure 5 in \citet{yama2014} shows the relation between the electron spectral parameters $p$ and $a$
on the relation of soft and hard X-ray spectral indices.

\begin{figure}
  \begin{center}
    \includegraphics[width=8cm]{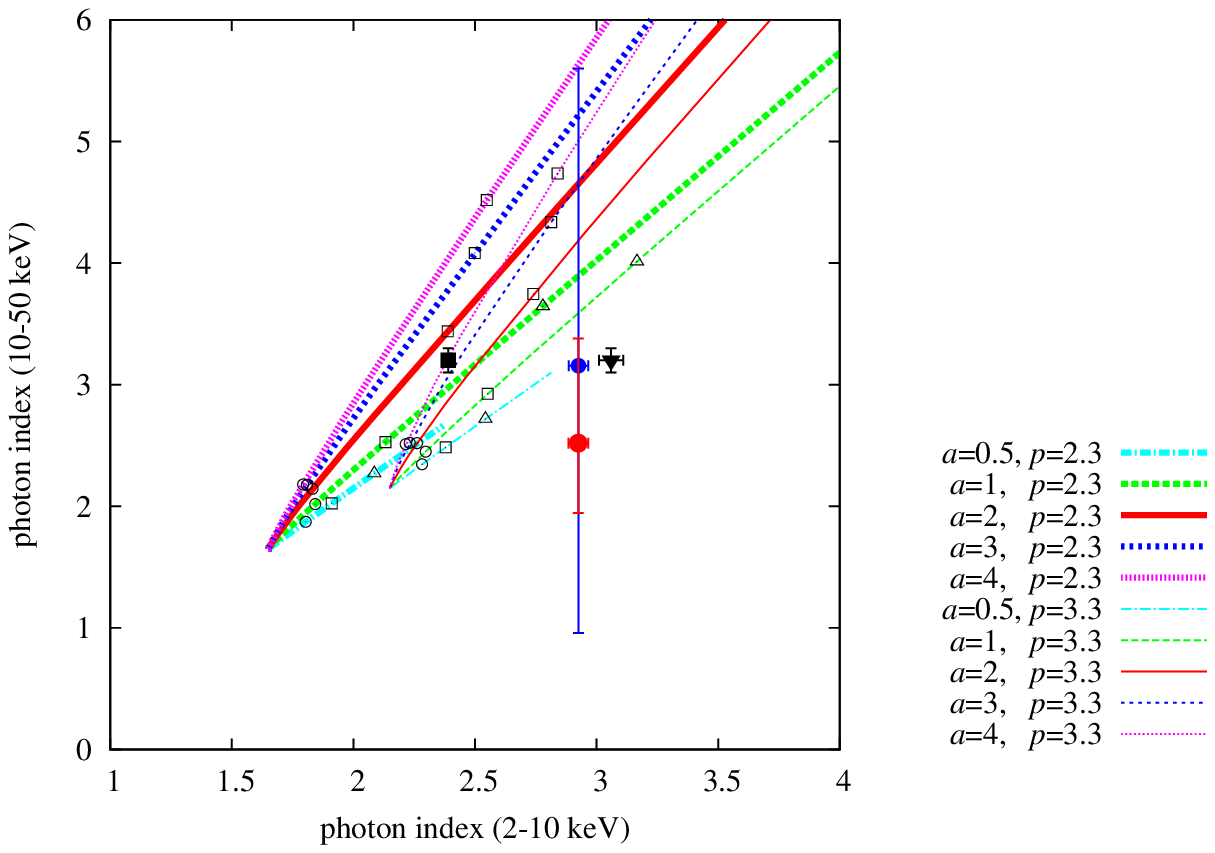}
  \end{center}
  \caption{The  relation between the two X-ray photon  indices for soft and hard bands for a breaking power-law model.
    The straight lines show the theoretical lines derived by \citet{yama2014}.
    The open squares, triangles and circles are for $BE_{\rm max,e}^{2} = 10^{4}$, $10^{5}$, and $10^{6} \mu{\rm G (TeV)}^{2}$,
    respectively (See \cite{yama2014}).
    The black squares and triangles show the result of RX J1713.7$-$3946 
    and  of Cassiopeia A, respectively.
    The red and blue circles show this result with model (iv) with cross-normalization fixed to 1.13 or freed, respectively. 
    These error bars show 99\% confidence level.
  }
  \label{fig:yama2014bekibeki}
\end{figure}

In order to discuss theoretical models,
we tried a fitting with a broken power-law model (iv) with the fixed breaking energy to 10 keV. 
Table \ref{tab:widebestfitpara} shows the best-fit parameters of this model,
and lower right panel of Figure \ref{fig:widespec} is the spectrum overlaid with the best-fit model spectra.
Both the derived photon  indices are consistent with that obtained  with the model (iii), though that in the hard band is marginally smaller by 0.1 than the latter.  The best-fit
flux  is close to that obtained in the  models (i) and (ii).
Figure \ref{fig:yama2014bekibeki}  overlays our results of the photon indices 
of below and above 10 keV  in the\ model (iv) in red on Figure 5 in \citet{yama2014}. 
We find that our data do not fit any of the theoretical-model lines by \citet{yama2014}.
Thus we find Vela Jr.\ is the second outlier of the theoretical lines following Cassiopeia A,
despite the spectrum in higher energy than rolloff.
They are unlike the other SNR RX J1713.7$-$3946 whose spectral shape is well described with 
power-law $+$ exponential cutoff model, exhibits the soft and hard photon indices 
on one of the theoretical lines.

A  possible cause for the mismatch is that 
at least one of the assumptions (a), (b)  and (c) is incorrect for this object.
A normal one-zone synchrotron X-ray spectrum  usually has a photon index of $\sim 2$  in a softer energy band 
and  rolls off toward the harder energy band due to cooling or escape.
However  the wide-band X-ray spectrum of Vela Jr.\ is  well reproduced with a single power-law or  even a concave-shape broken power-law.
It may suggest that combination of  more than one emission component and/or  complex emission mechanisms  creates the observed X-ray spectra
(\cite{long1994}, \cite{drur1999}, \cite{zira2007}, \cite{malk2001}, \cite{topt1987}, \cite{medv2000}, \cite{revi2010}, \cite{tera2011},
\cite{yama2006}, \cite{lami2001}, \cite{vink2003}, \cite{vink2008}, \cite{ohir2012}).

 In addition, in order to validate the cross-normalization between the XIS and the PIN,
we fit the spectra, allowing both the XIS and PIN normalizations to vary independently, 
and find that the best-fit photon index and the XIS flux are consistent with the above-discussed case within the error range, whereas
the derived cross-normalization factor is $1.4 \pm 0.3$, 
which is  marginally larger than that reported for the point source.
This result may imply that  brightness distribution in hard X-rays is  more compact than that of the XIS image
and that we have actually underestimated the PIN effective area, as we have  assumed  a larger diffuse-emission region than the real one.
Considering the case of  more compact hard X-ray brightness distribution than that of soft X-rays,
we also showed a conservative fitting result with freed cross-normalization factor in 
figure \ref{fig:yama2014bekibeki} with blue mark and error bars.
The best-fit photon indices are $2.93 \pm 0.04$ and $3.16_{-2.20}^{+2.44}$ in the soft and hard energy range, respectively.
Although this estimation accepts most of the model lines presented in \citet{yama2014}, 
it requires  more than one emission region or  complex emission mechanisms. 
 
Lastly, we show flux comparison with TeV emission.
In section \ref{sec:xisspec}, we derived the flux 
of $(4.43 \pm 0.03) \times 10^{-11} {\rm erg\ cm}^{-2}\ {\rm s}^{-1}$ in 2--10 keV band
while \citet{ahar2007} showed the TeV flux of entire Vela Jr.\ 
of $(15.2 \pm 0.7 \pm 3.20) \pm 10^{-12} {\rm cm}^{-2} {\rm s}^{-1}$ with H.E.S.S.
Both band spectra exhibit similar slope and the X-ray to TeV gamma-ray flux ratio is $\sim$ 2.91.
If we assume the cosmic microwave background inverse Compton scattering as TeV emission mechanism,
we estimate the magnetic field $B \sim 5.5\ \mu$G, which is consistent value 
derived by \citet{kish2013} and \citet{lee2013}.
If the TeV emission is hadronic, the field strength is not constrained 
and it may be much higher \citep{bamb2005b}.


\begin{ack}
The authors would like to thank all the members of the Suzaku team 
for their continuous contributions in the maintenance of onboard instruments,
spacecraft operation, calibrations, software development, and user support both in Japan and the United States.
We would like to thank associate professor Koji Mori for analysis of the XIS image.
We would like to thank Prof. Yasushi Fukazawa and Dr. Takaaki Tanaka for useful comments 
about analysis of HXD-PIN. 
This work was supported in part
by Grants-in-Aid for Scientific Research from 
the Ministry of Education, Culture, Sports, Science and Technology (MEXT)
(No.~23340055, Y.~T, No.~15K05088, R.~Y, No.~ 15K05107, A.~B., No.~15H03642-01, M.~S.~T., and No.~25800119, S. K.).
\end{ack}

\appendix




\end{document}